\documentclass[preprint,prd,aps,nofootinbib]{revtex4}
\usepackage{graphicx}
\usepackage{dcolumn}
\usepackage{bm}
\usepackage{graphicx}
\usepackage{diagbox}
\usepackage{slashed}
\usepackage{float}
\usepackage{color}
\usepackage{color}
\usepackage[dvipsnames]{xcolor}
\usepackage{amssymb}
\usepackage{float}
\usepackage{subfigure}

\definecolor{black-blue}{RGB}{77,116,175}
\definecolor{black-yellow}{RGB}{231,162,33}
\definecolor{black-green}{RGB}{144,180,58}
\definecolor{black-red}{RGB}{246,95,50}
\begin{document}
\title{Relativistic effects in the strong and electromagnetic decays of the ${D^*}$ meson}
\author{Man Jia$^{1,2,3}$\footnote{manjia1@126.com},
 Wei Li$^{4}$,
 Su-Yan Pei$^{1,2,3}$,
 Xin-Yao Du$^{1,2,3}$,
 Guo-Zhu Ning$^{1,2,3}$,
 Guo-Li Wang$^{1,2,3}$\footnote{wgl@hbu.edu.cn, corresponding author}}

\affiliation{$^1$ Department of Physics, Hebei University, Baoding 071002, China
\nonumber\\
$^{2}$ Hebei Key Laboratory of High-precision Computation and Application of Quantum Field Theory, Baoding 071002, China
\nonumber\\
$^{3}$ Hebei Research Center of the Basic Discipline for Computational Physics, Baoding 071002, China
\nonumber\\
$^4$ College of Science, Hebei Agriculture University, Baoding 071001, China
\nonumber\\}
\begin{abstract}
In this paper, we solve the complete Salpeter equation and use the obtained relativistic wave function to calculate the strong and radiative electromagnetic decays of the ${D^*}$ meson. { We obtain the results $\Gamma(D^{*}(2007)^{0}\to D^{0}\pi^{0})=34.6~\rm{keV}$ and $\Gamma(D^{*}(2007)^{0}\rightarrow D^{0}\gamma)=19.4~\rm{keV}$, and the estimated full width is $\Gamma(D^{*}(2007)^{0})=54.0~\rm{keV}$.} The focus of this study is on the relativistic corrections. In our method, the wave function of the $D$ meson is not a pure $S$-wave, but includes both a non-relativistic $S$-wave and a relativistic $P$-wave, while the wave function of the $D^*$ meson includes a non-relativistic $S$-wave as well as both relativistic $P$-wave and $D$-wave. Therefore, in this case, the decay ${D^{*}\rightarrow{D}\gamma}$ is not a non-relativistic $M1$ transition, but rather an $M1+E2+M3+E4$ decay. We find that in a strong decay $D^{*}\rightarrow{D}{\pi}$, the non-relativistic contribution is dominant, while in an electromagnetic decay ${D^{*}\rightarrow{D}\gamma}$, the relativistic correction is dominant.

\end{abstract}
\maketitle
\section{Introduction}
The $D^{*}$ mesons were discovered many years ago \cite{1976,1977,slac-pub-1973(1)}, and since then, they have received considerable research attention. Many experiments have studied their properties, as reported in Refs. \cite{SLAC-PUB-2916,JADE Collaboration,SLAC-PUB-4518,0000089,HRS,butler1992,bartelt1998,albrecht1995,ppe-92-017}. The mass of the $D^{*}$ meson has been accurately measured in various studies. For example, the mass of ${D^*}{(2007)^0}$ was precisely determined as $M=2006.85\pm0.05$~MeV, and the mass of ${D^*}{(2010)^+}$ as $M=2010.26\pm0.05$~MeV \cite{PDG}. However, the measurement uncertainty of the decay branch ratios is quite substantial. Reported values include $Br({D^{*0}\rightarrow{D^0}{\pi^0}})=64.7\pm0.9\%$, $Br({D^{*0}\rightarrow{D^0\gamma}})=35.3\pm0.9\%$, $Br({D^{*+}\rightarrow{D^0}{\pi^+}})=67.7\pm0.5\%$, $Br({D^{*+}\rightarrow{D^0}{\pi^0}})=30.7\pm0.5\%$, and  $Br({D^{*+}\rightarrow{D^+}\gamma})=1.6\pm0.4\%$ \cite{PDG}.  The uncertainty of the total width of ${D^*}{(2010)^+}$, $\Gamma({D^*}{(2010)^+})=83.4\pm1.8$~keV, is still relatively large, while the full width of ${D^*}{(2007)^0}$ has not been measured and only has an upper limit, $\Gamma({D^*}{(2007)^0})<2.1$~MeV \cite{PDG}.

Many theoretical studies have investigated the properties of ${D^*}{(2010)^+}$ and ${D^*}{(2007)^0}$ \cite{Print-88-0167,eichten1980,pham1982,SLAC-PUB-3522,9203137,9209239v1,9610412v1(1)}. For example, in Ref.\cite{9209239v1}, heavy hadron chiral perturbation theory was chosen and experimental data were used as input to study the strong and $M1$ decays of $D^*$. In Ref.\cite{9209262v1(1)}, heavy quark symmetry and chiral symmetry were used to study the radiative decay, while Ref.\cite{9406300v1(1)} used the relativistic light-front quark model to calculate the strong and radiative decays. Quantum chromodynamics (QCD) sum rules based on light-cone expansion were used in Ref.\cite{aliev1996} to estimate the transition amplitude and decay rates, and QCD sum rules were also used in Ref.\cite{9610412v1(1)} to analyze the radiative decay. In
Ref.\cite{eichten}, the chiral quark model with $1/m_c$ corrections was chosen, and experimental data were used as input to calculate the radiative decay, while Ref. \cite{close2005} used the $^{3}P_{0}$ model to calculate the strong decay. However, as will be shown later (Tables \ref{I} and \ref{IV}), different theoretical models result in significant differences in the results obtained, indicating the complexity of theoretical calculations and thus the need for more careful theoretical study.

Therefore, in this paper, we add a calculation in the framework of the instantaneous Bethe-Salpeter (BS) equation method, with a focus on relativistic corrections. We study the Okubo-Zweig-Iizuka (OZI) rule allowing strong decay and radiative electromagnetic (EM) decays of the ${D^*}{(2007)^0}$ and ${D^*}{(2010)^+}$ mesons. The BS equation was proposed by Bethe and Salpeter in 1951 \cite{salpeter1951}. However, given the challenges associated with solving this complex and difficult equation, in 1952 Salpeter introduced the instantaneous approximation version of the BS equation \cite{salpeter1952}, which is particular suitable for heavy mesons. We previously applied the solutions of the Salpeter equation to calculate the strong decays \cite{1305.1067v2,tan2018,wang-2018} and radiative EM transition processes \cite{main,wang2013,PhysRevD.108.033003} of heavy mesons and found that this method could obtain results consistent with experimental data. Therefore, in this article, we adopt this method and focus on the  relativistic corrections. For example, for the ${D^{*}\rightarrow{D}\gamma}$ decay, our result is not the $M1$ transition, but the $M1+E2+M3+E4$ decay.

The remainder of the paper is organized as follows. In Sect.2, we show our method for calculating the transition amplitudes for the strong and EM decays and the relativistic wave functions used in this paper. In Sect.3, we present and discuss our results. Finally, a brief summary is given in Sect.4.

\section{ The theoretical calculations }

\subsection{Transition amplitude of strong decay}

The ${D^*}{(2007)^0}$ and ${D^*}{(2010)^+}$ particles are allowed strong decay by the Okubo-Zweig-Iizuka(OZI) rule. We take ${D^*}{(2007)^0}\rightarrow{D^0}{\pi^0}$ as an example of our method for calculating the strong transition amplitude. Due to the light mass of the $\pi$ meson, the instantaneous approximation is not good for $\pi$, so we aim to avoid using the wave function of $\pi$. We adopt a reduction formula, then the transition S-matrix for the decay ${D^*}{(2007)^0}\rightarrow{D^0}{\pi^0}$ can be written as
\begin{eqnarray}\label{am1}
&\langle{D^0}({P_{f1}}){\pi^0}(P_{f2})|{D^{*0}}(P)\rangle\nonumber\\&\quad
=\int{d^4}xe^{iP_{f_2}\cdot x}(M^{2}_{\pi}\nonumber\\&\quad-P^{2}_{f2})
\langle{D^0}({P_{f1}})|\Phi_{\pi}(x)|{D^{*0}}(P)\rangle,\nonumber\\
\end{eqnarray}
\begin{figure}[!htb]
\centering
\includegraphics[width=3.9in]{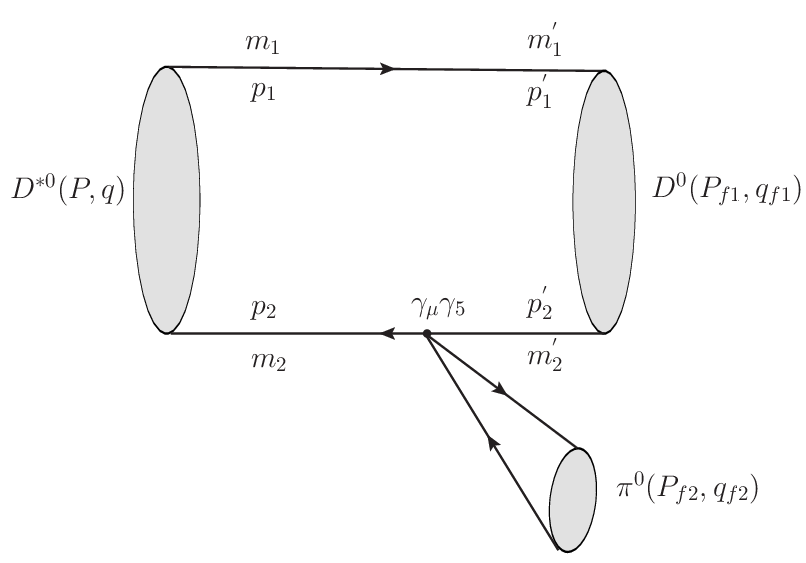}
\caption{The Feynman diagram for the two-body strong decay.}
\label{SDFeymp}
\end{figure}
\begin{figure}[!htb]
\begin{minipage}[c]{1\textwidth}
\includegraphics[width=3in]{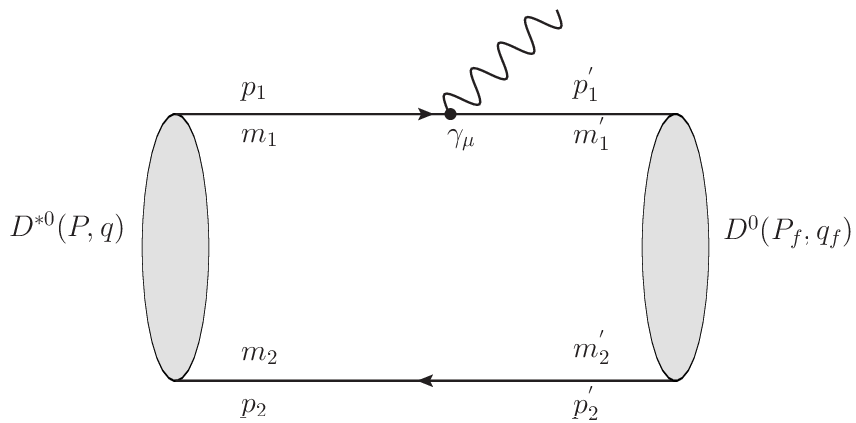}
\includegraphics[width=3in]{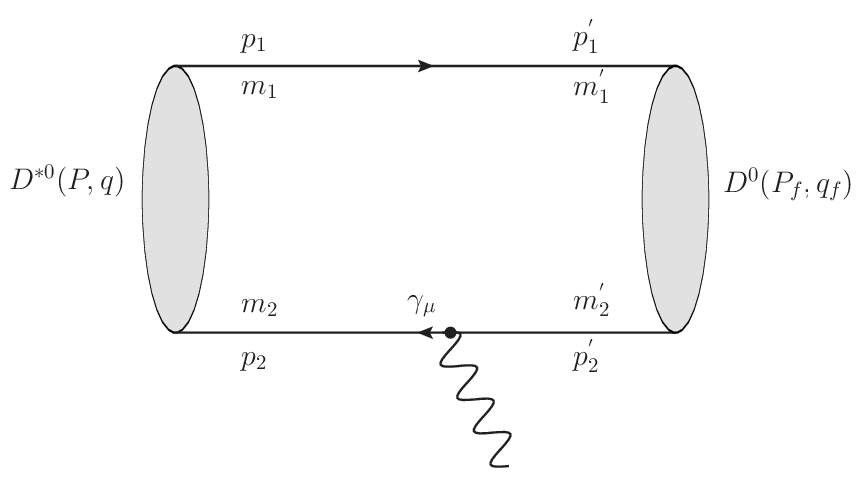}
\end{minipage}
\caption{The Feynman diagram for the EM decay.}
\label{feynman}
\end{figure}
where $P$, $P_{f1}$, and $P_{f2}$ are the momenta of the $D^{*0}$, $D^{0}$, and $\pi^{0}$ mesons, respectively; $M_{\pi}$ is the mass of $\pi^{0}$; and $\Phi_{\pi}(x)$ is the $\pi$ meson field. Using the partial pseudovector current conservation (PCAC) rule in field theory and the low energy theorem, Eq. (\ref{am1}) becomes
\begin{eqnarray}\label{am2}
-\frac{iP^{\mu}_{f2}}{f_{\pi}}(2\pi)^{4}\delta^{4}(P-P_{f1}-P_{f2})
\langle{D^0}({P_{f1}})|\bar{u}\gamma_{\mu}\gamma_{5}u|{D^{*0}}(P)\rangle,
\end{eqnarray}
where $f_{\pi}$ is the decay constant of $\pi$. In this case, the corresponding Feynman diagram is illustrated in Fig. \ref{SDFeymp}.

According to the Mandelstam formula \cite{mandelstam1955}, the transition amplitude in our method can be written as \cite{0505205v2}
\begin{eqnarray}\label{am4}
&&\mathcal{M}=\langle{D^0}({P_{f1}})|\bar{u}\gamma_{\mu}\gamma_{5}u|{D^{*0}}(P)\rangle=
\int\frac{d^{3}q}{(2\pi)^{3}}Tr\left[\bar{\varphi}^{++}_{P_{f1}}(q_{_{f\bot}})\gamma_{\mu}\gamma_{5}\varphi^{++}_{P}(q_{_{\bot}})\frac{\slashed{P}}{M}\right],
\end{eqnarray}
where $q_{_\bot}=q-(P\cdot q/M^2)P$ ($q_{_\bot}=(0,\vec{q})$ in the center-of-mass system of the initial state) is the instantaneous relative momentum between quarks and antiquarks, $M$ is the mass of $D^{*0}$, and $\bar{\varphi}^{++}_{P_{_{f1}}}(q_{_{f\bot}})$ and $\varphi^{++}_{_{P}}(q_{_{\bot}})$ are the positive energy BS wave functions of $D^{0}$ and $D^{*0}$, respectively.

\subsection{Transition amplitude of EM decay}

The Feynman diagram for the EM decay $D^{*0}\rightarrow D^0+\gamma$ is shown in Fig. \ref{feynman}. The S-matrix element of the process can be written as
\begin{eqnarray}\label{am5}
&&\langle{D^0}({P_{f}})\gamma(\kappa,\epsilon_{0})|{D^{*0}}(P,\epsilon)\rangle=(2\pi)^{4}\delta^{4}(P-P_{f}-\kappa)\epsilon_{0\mu}\mathcal{M^{\mu}},
\end{eqnarray}
where $\epsilon$ and $\epsilon_{0}$ are the polarization vectors of $D^{*0}$ and photon, respectively; and $P$, $P_{f}$, and $\kappa$ represent the momenta of $D^{*0}$, $D^{0}$, and a photon, respectively.

We can see that the invariant amplitude $\mathcal{M^{\mu}}$ consists of two parts, where photons are emitted from quarks and antiquarks, respectively. Using the wave functions obtained by solving the complete Salpeter equation \cite{salpeter1952,mainext}, the transition amplitude can be written as \cite{chao-hsi2006}
\begin{eqnarray}\label{am9}
\mathcal{M^{\mu}}=\int\frac{d^{3}q}{(2\pi)^{3}}
Tr\left[Q_{1}\frac{\slashed{P}}{M}\overline{\varphi}_{P_{f}}^{++}(q_{_{\perp}}+\alpha_{2}P_{_{f\perp}})\gamma^{\mu}\varphi_{P}^{++}(q_{_{\perp}})
+Q_{2}\overline{\varphi}_{P_{f}}^{++}(q_{_{\perp}}-\alpha_{1}P_{_{f\perp}})\frac{\slashed{P}}{M}\varphi_{P}^{++}(q_{_{\perp}})\gamma^{\mu}\right],
\end{eqnarray}
where $\overline{\varphi}_{P_{f}}^{++}=\gamma^0 {{\varphi}_{P_{f}}^{++}}^{\dagger}\gamma_0$; $Q_{1}$ and $Q_{2}$ (in units of $e$) are the charges of quarks and antiquarks, respectively; $\alpha_{1}=\frac{m_{1}}{m_{1}+m_{2}}$ and $\alpha_{2}=\frac{m_{2}}{m_{1}+m_{2}}$, and $m_{1}$ and $m_{2}$ are the masses of the quarks and  antiquarks, respectively.

\subsection{The relativistic wave functions}

In the instantaneous approximation, the general relativistic wave function for a $1^{-}$ meson ${D^*}{(2007)^0}$ or ${D^*}{(2010)^+}$ is written as \cite{0512009v1}
\begin{eqnarray}\label{am10}
\varphi_{1^{-}}(q_{_{\perp}})=&&q_{_{\perp}}\cdot\epsilon\left[b_{1}+\frac{\slashed{P}}{M}b_{2}
+\frac{\slashed{q}_{_{\perp}}}{M}b_{3}+\frac{\slashed{P}\slashed{q}_{_{\perp}}}{M^{2}}b_{4}\right]
+M\slashed{\epsilon}b_{5}+\slashed{\epsilon}\slashed{P}b_{6}
\nonumber\\&&+(\slashed{q}_{_{\perp}}\slashed{\epsilon}-q_{_{\perp}}\cdot\epsilon)b_{7}
+\frac{1}{M}(\slashed{P}\slashed{\epsilon}\slashed{q}_{_{\perp}}-\slashed{P}q_{_{\perp}}\cdot\epsilon)b_{8},
\end{eqnarray}
where the radial wave function $b_{i}~(i=1,2,\ldots,8)$ is a function of $-q_{_{\perp}}^{2}$ ($\equiv\vec{q}^2$ in the center-of-mass system of ${D^*}$).
{Now we show that each term in this $1^{-}$ wave function has negative parity $\eta_{_P}=-1$ .
When we perform the parity transformation, $P'=(P_0,-\vec{P})$, $q'=(q_0,-\vec{q})$, and the wave function becomes
{$$\varphi^{\prime}_{_{1^{-}}}(P^{\prime},q^{\prime}_{_\perp})=\gamma_{_0}\varphi_{_{1^{-}}}(P',q{'}_{_\perp})\gamma_{_0}=\eta_{_P}\varphi_{_{1^{-}}}(P,q_{_\perp}).$$}
{In the center-of-mass system (CMS) of momentum $P$}, {we have $P'=(M,0)=P$, $q{'}_{_\perp}=(0,-\vec{q})=-q_{_\perp}$}, and
\begin{eqnarray}\label{parity}
\gamma_{_0}\varphi{'}_{_{1^{-}}}(q{'}_{_\perp})\gamma_{_0}=
&&\gamma_{_0}\bigg\{-q_{_{\perp}}\cdot\epsilon\left[b_{1}+\frac{\slashed{P}}{M}b_{2}
-\frac{\slashed{q}_{_{\perp}}}{M}b_{3}-\frac{\slashed{P}\slashed{q}_{_{\perp}}}{M^{2}}b_{4}\right]
+M\slashed{\epsilon}b_{5}+\slashed{\epsilon}\slashed{P}b_{6}
\nonumber\\&&-(\slashed{q}_{_{\perp}}\slashed{\epsilon}-q_{_{\perp}}\cdot\epsilon)b_{7}
-\frac{1}{M}(\slashed{P}\slashed{\epsilon}\slashed{q}_{_{\perp}}-\slashed{P}q_{_{\perp}}\cdot\epsilon)b_{8}\bigg\}\gamma_{_0}
\nonumber\\&&=-\varphi_{_{1^{-}}}(P,q_{_\perp}),
\end{eqnarray}
where we have $\epsilon=(0,\vec {\epsilon})$, since $P\cdot \epsilon=0$ in the CMS of the meson, and under the parity transformation, $\epsilon^{\prime}=\epsilon$ and $\gamma_0\slashed{\epsilon}^{\prime}\gamma_0=-\slashed{\epsilon}$. Equation(\ref{parity}) shows us that the parity of the wave function is $\eta_{_P}=-1$.

Further, we show that the $1^{-}$ wave function contains an $S$-wave, $P$-wave, and $D$-wave simultaneously, indicating that this $1^{-}$ particle is in an $S-P-D$ mixing state. We rewrite the wave function in terms of the spherical harmonics $Y_{_{lm}}$, for example
$$
\slashed{\epsilon}=\sqrt{4\pi}Y_{00}(\epsilon^{+}\gamma^{-}+\epsilon^{-}\gamma^{+}-\epsilon^{\Delta}\gamma^{\Delta}),
$$
where $\gamma^{+}=-\frac{\gamma^{1}+i\gamma^{2}}{\sqrt{2}}$, $\gamma^{-}=\frac{\gamma^{1}-i\gamma^{2}}{\sqrt{2}}$ and $\gamma^{\Delta}=\gamma^{3}$; $\epsilon^{+}$, $\epsilon^{-}$, and $\epsilon^{\Delta}$ have similar definitions.
It is easy to see that the $b_{_5}$ and $b_{_6}$ terms are $S$-waves. Since we have
$$
q_{_\perp}\cdot\epsilon=\sqrt{\frac{4\pi}{3}}|\vec{q}|
\left(-Y_{11}\epsilon^{-}-Y_{1-1}\epsilon^{+}
+Y_{10}\epsilon^{\Delta}\right),
$$
and
$$
\slashed{q}_{_\perp}\slashed{\epsilon}=-\sqrt{\frac{4\pi}{3}}|\vec{q}|
\left(-Y_{11}\gamma^{-}-Y_{1-1}\gamma^{+}
+Y_{10}\gamma^{\Delta}\right)\left(\epsilon^{\Delta}\gamma^{\Delta}-\epsilon^{-}\gamma^{+}-\epsilon^{+}\gamma^{-}\right),
$$
the $b_{_1}$, $b_{_2}$, $b_{_7}$, and $b_{_8}$ terms are $P$-waves. For the $b_{_3}$ and $b_{_4}$ terms, we have
$$
q_{_\perp}\cdot\epsilon\slashed{q}_{_\perp}=
\frac{1}{3}{q}^2_{_\perp}+\left[(q_{_\perp}\cdot\epsilon\slashed{q}_{_\perp})
-\frac{1}{3}{q}^2_{_\perp}\right]
$$$$=\frac{1}{3}|\vec{q}|^{2}
\left(\epsilon^{\Delta}\gamma^{\Delta}-\epsilon^{-}\gamma^{+}-\epsilon^{+}\gamma^{-}\right)+\frac{2}{3}\sqrt{\frac{\pi}{5}}
|\vec{q}|^{2}\left[\sqrt{6}\epsilon^{+}\gamma^{+}Y_{2-2}-\sqrt{3}\left(\epsilon^{+}\gamma^{\Delta}+\epsilon^{\Delta}\gamma^{+}\right)Y_{2-1}\right.
$$$$\left.+\left(2\epsilon^{\Delta}\gamma^{\Delta}+\epsilon^{-}\gamma^{+}+\epsilon^{+}\gamma^{-}\right)Y_{20}-\sqrt{3}\left(\epsilon^{-}\gamma^{\Delta}
+\epsilon^{\Delta}\gamma^{-}\right)Y_{21}+\sqrt{6}\epsilon^{-}\gamma^{-}Y_{22}\right],
$$
so $b_{_3}$ and $b_{_4}$ terms are an $S$-$D$ mixture. In Eq.(\ref{am10}), the complete $S$-wave is
\begin{eqnarray}
&&M\slashed{\epsilon}(b_{5}+\frac{\slashed{P}}{M}b_{6})+\frac{1}{3}{q}^2_{_\perp}\slashed{\epsilon}(\frac{1}{M}b_{3}
-\frac{\slashed{P}}{M^{2}}b_{4}),
\end{eqnarray}
and the pure $D$-wave is
\begin{eqnarray}
&&\left(q_{\perp}\cdot\epsilon~\slashed{q}_{\perp}-\frac{1}{3}{q}^2_{_\perp}\slashed{\epsilon}\right)
\left(\frac{1}{M}b_{3}-\frac{\slashed{P}}{M^{2}}b_{4}\right).
\end{eqnarray}
Therefore, the widely used representations of $P=(-1)^{L+1}$ and $^{2\,S+1}L_J$ are strictly valid only in non-relativistic conditions and are not applicable to relativistic cases \cite{JHEP05(2022)006,pei}.}

The corresponding positive wave function can be written as
\begin{eqnarray}\label{am11}
\varphi_{1^{-}}^{++}(q_{_{\perp}})=q_{_{\perp}}\cdot\epsilon\left(B_{1}+B_{2}\frac{\slashed{P}}{M}+B_{3}\frac{\slashed{q}}{M}
+B_{4}\frac{\slashed{P}\slashed{q}_{_{\perp}}}{M^{2}}\right)
\nonumber\\+M\slashed{\epsilon}\left(B_{5}+B_{6}\frac{\slashed{P}}{M}+B_{7}\frac{\slashed{q}_{_{\perp}}}{M}
+B_{8}\frac{\slashed{P}\slashed{q}_{_{\perp}}}{M^{2}}\right),
\end{eqnarray}
where $B_{i}(i=1,2,\ldots,8)$ is related to the independent radial wave functions $b_{3}$, $b_{4}$, $b_{5}$, and $b_{6}$, whose numerical values are obtained by solving the full Salpeter equation \cite{0512009v1},  resulting in the following relations.
$$
B_{1}=\frac{1}{2M(m_{1}\omega_{2}+m_{2}\omega_{1})}\left[-(\omega_{1}+\omega_{2})\vec{q}^{2}b_{3}\right.
$$
$$~~~~~~\left.-(m_{1}+m_{2})\vec{q}^{2}b_{4}+2M^{2}\omega_{2}b_{5}-2M^{2}m_{2}b_{6}\right],
$$
$$
B_{2}=\frac{1}{2M(m_{1}\omega_{2}+m_{2}\omega_{1})}\left[-(m_{1}-m_{2})\vec{q}^{2}b_{3}\right.
$$
$$~~~~~~\left.-(\omega_{1}-\omega_{2})\vec{q}^{2}b_{4}-2M^{2}m_{2}b_{5}+2M^{2}\omega_{2}b_{6}\right],
$$
$$
B_{3}=\frac{1}{2}\left[b_{3}+\frac{(m_{1}+m_{2})}{\omega_{1}+\omega_{2}}b_{4}-\frac{2M^{2}}{m_{1}\omega_{2}+m_{2}\omega_{1}}b_{6}\right],~~~
B_{5}=\frac{1}{2}\left[b_{5}-\frac{\omega_{1}+\omega_{2}}{m_{1}+m_{2}}b_{6}\right],
$$
$$
B_{4}=\frac{1}{2}\left[\frac{\omega_{1}+\omega_{2}}{m_{1}+m_{2}}b_{3}+b_{4}-\frac{2M^{2}}{m_{1}\omega_{2}+m_{2}\omega_{1}}b_{5}\right],~~~B_{6}=\frac{1}{2}\left[-\frac{(m_{1}+m_{2})}{\omega_{1}+\omega_{2}}b_{5}+b_{6}\right],
$$
$$
B_{7}={M}B_5\frac{\omega_{1}-\omega_{2}}{m_{1}\omega_{2}+m_{2}\omega_{1}},~~~
B_{8}=-{M}B_5\frac{m_{1}+m_{2}}{m_{1}\omega_{2}+m_{2}\omega_{1}},
$$
where $\omega_{i}=\sqrt{m_i^{2}-q_{_{\perp}}^{2}}$ ($i=1,2$) is the energy of quarks or antiquarks. We note that in Eq.(\ref{am11}), although each term has a quantum number of $1^-$, not every term is an $S$-wave. Among them, the $B_{5}$ and $B_{6}$ terms are $S$-waves, the $B_{1}$, $B_{2}$, $B_{7},$ and $B_{8}$ terms are $P$-waves, and the $B_{3}$ and $B_{4}$ terms are a mixture of $S$-waves and $D$-waves \cite{PhysRevD.108.033003}.

The relativistic wave function of the $0^{-}$ state is \cite{mainext}
\begin{eqnarray}\label{am12}
\varphi_{0^{-}}(q_{_{f\perp}})=\left(\slashed{P}_{f}a_{1}+M_{f}a_{2}+\slashed{q}_{f\perp}a_{3}
+\frac{\slashed{P}_{f}\slashed{q}_{f\perp}}{M_{f}}a_{4}\right)\gamma_{5}.
\end{eqnarray}
Its corresponding positive energy wave function is
\begin{eqnarray}\label{am13}
\varphi_{0^{-}}^{++}(q_{_{f\perp}})=\left(\frac{\slashed{P}_{f}}{M_{f}}A_{1}+A_{2}+\frac{\slashed{q}_{f\perp}}{M_{f}}A_{3}
+\frac{\slashed{q}_{f\perp}\slashed{P}_{f}}{M_{f}^{2}}A_{4}\right)\gamma_{5},
\end{eqnarray}
where $A_{i}(i=1,2,3,4)$ is related to the independent radial wave function $a_{1}$ and $a_{2}$
$$A_{1}=\frac{M_{f}}{2}\left[a_{1}+\frac{m'_{1}+m'_{2}}{\omega'_{1}+\omega'_{2}}a_{2}\right],$$
$$A_{2}=\frac{M_{f}}{2}\left[\frac{\omega'_{1}+\omega'_{2}}{m'_{1}+m'_{2}}a_{1}+a_{2}\right],$$
$$A_{3}=-\frac{M_{f}(\omega'_{1}-\omega'_{2})}{m'_{1}\omega'_{2}+m'_{2}\omega'_{1}}A_{2},~~~~~~A_{4}=-\frac{M_{f}(m'_{1}+m'_{2})}{m'_{1}\omega'_{2}+m'_{2}\omega'_{1}}A_{2}.$$
Similarly, the $A_{1}$ and $A_{2}$ terms are $S$-waves, and the $A_{3}$ and $A_{4}$ terms are $P$-waves \cite{JHEP05(2022)006}.

\section{Results and disscussion}
In our numerical calculation, the interaction kernel is composed of a linear potential $V_s$ and a vector potential $V_v$ \cite{mainext,chang2010},
$$
V_{s}(\vec{q})=-\left(\frac{\lambda}{\alpha}+V_{0}\right)\delta^{3}(\vec{q})+\frac{\lambda}{\pi^{2}}
\frac{1}{(\vec{q}^2+\alpha^{2})^{2}},
$$
$$
V_{v}(\vec{q})=-\frac{2}{3\pi^{2}}\frac{\alpha_{s}(\vec{q})}{(\vec{q}^2+\alpha^{2})},
$$ where $V_0$ is a free parameter to fit the meson spectroscopy, and
$$
\alpha_{s}(\vec{q})=\frac{12\pi}{33-2N_{f}}\frac{1}{\log\bigg(a+\frac{\vec{q}^{2}}{\Lambda^{2}_{QCD}}\bigg)},
$$ is the running coupling constant.
Therefore, the model-dependent parameters $\lambda=0.15~\rm{GeV}^{2}$, $\Lambda_{QCD}=0.18~\rm{GeV}$, $a=e=2.7183$, and $N_f=3$, and the quark masses $m_{_u}=0.374~\rm{GeV}$, $m_{_d}=0.38~\rm{GeV}$, and $m_{_c}=1.62~\rm{GeV}$ are usde.
\subsection{Strong decay widths of ${D^*}{(2007)^0}$ and ${D^*}{(2010)^+}$}
With the upper parameters, the calculated strong decay widths are
\begin{eqnarray}\label{am26}
\Gamma(D^{*}(2007)^{0}\rightarrow D^{0}\pi^{0})={34.6~\rm{keV}},
\end{eqnarray}
\begin{eqnarray}\label{am27}
\Gamma(D^{*}(2010)^{+}\rightarrow D^{0}\pi^{+})={55.8~\rm{keV}},
\end{eqnarray}
\begin{eqnarray}\label{am28}
\Gamma(D^{*}(2010)^{+}\rightarrow D^{+}\pi^{0})={25.8~\rm{keV}}.
\end{eqnarray}
\begin{table}[!htb]
\caption{ The strong decay widths (keV) of ${D^*}{(2007)^0}$ and ${D^*}{(2010)^+}$.}
\begin{tabular}{cccccccccc}
\hline
& ~~{\cite{9610412v1(1)}}~~ & ~~{\cite{close2005}}~~  & ~~{\cite{aliev1996}}~~ & ~~{\cite{cheung2014}}~~ & ~~{\cite{jaus1996}} ~~ & ~~{\cite{eichten}}~~ & ~~Ours ~~ &~~Ex\cite{PDG} ~~\\
\hline\hline
$\Gamma(D^{*}(2010)^{+}\rightarrow D^{0}\pi^{+})$ & & $ 25 $ &   $ 32\pm5 $&56.6&62.53& $ 65 $ & $ {55.8} $ & $ 56.5\pm1.6 $ \\
 \hline
$\Gamma(D^{*}(2010)^{+}\rightarrow D^{+}\pi^{0})$ & & $ 11 $ &   $ 15\pm2 $&25.8&28.30& $ 30 $ & $ {25.8} $ & $ 25.6\pm1.0 $ \\
 \hline
 $\Gamma(D^{*}(2007)^{0}\rightarrow D^{0}\pi^{0})$ &$23\pm3$ & $16$&$22\pm2$&36.9&43.40& 42& $ {34.6} $ &
\\\hline\hline
$\frac{\Gamma(D^{*}(2010)^{+}\rightarrow D^{0}\pi^{+})}{\Gamma(D^{*}(2010)^{+}\rightarrow D^{+}\pi^{0})}$& &2.27& 2.13&2.19&2.21 &2.17&{2.16} &2.21
\\\hline
$\frac{\Gamma(D^{*}(2010)^{+}\rightarrow D^{0}\pi^{+})}{\Gamma(D^{*}(2007)^{0}\rightarrow D^{0}\pi^{0})}$& &1.56& 1.45&1.53&1.44 &1.55&{1.61} &
\\\hline\hline
\end{tabular}
\label{I}
\end{table}
We list our results and other theoretical results as well as the available experimental data in Table \ref{I} for comparison. {Our results are very close to those of Ref.\cite{cheung2014} and the experimental data \cite{PDG}, and larger than the results in Refs. \cite{9610412v1(1),close2005,aliev1996}, but smaller than those of Refs. \cite{jaus1996,eichten}}.

For the particle $D^{*}(2007)^{0}$, there is currently a lack of precise experimental data regarding the total decay width of $D^{*}(2007)^{0}$.  The experiment only reported an upper limit of 2.1 \rm{MeV} \cite{0000089},  
and the Particle Date Group (PDG) provides branching ratios of $D^{*}(2007)^{0}\rightarrow D^{0}\pi^{0}$ and $D^{*}(2007)^{0}\rightarrow D^{0}\gamma$. Some studies have theoretically investigated the decay properties of $D^{*}(2007)^{0}$, such as the predictions of strong decay presented in Table \ref{I}. As can be seen from the table, there are some differences between the results of different models. To eliminate model dependence, we provide ratios $\frac{\Gamma(D^{*}(2010)^{+}\rightarrow D^{0}\pi^{+})}{\Gamma(D^{*}(2010)^{+}\rightarrow D^{+}\pi^{0})}$ and $\frac{\Gamma(D^{*}(2010)^{+}\rightarrow D^{0}\pi^{+})}{\Gamma(D^{*}(2007)^{0}\rightarrow D^{0}\pi^{0})}$. It can be seen that all the theoretical ratios are consistent with each other, and theoretical ratio $\frac{\Gamma(D^{*}(2010)^{+}\rightarrow D^{0}\pi^{+})}{\Gamma(D^{*}(2010)^{+}\rightarrow D^{+}\pi^{0})}$ is also consistent with the experimental data. Therefore, we can conclude that these ratios greatly reduce the model dependence.

In this article, we mainly focus on relativistic corrections in calculations; therefore, we provide the contributions of different partial waves to the width and show the results in Tables \ref{II} and \ref{III}. Here, $``complete"$ means that the complete wave function is used, while $``S~wave"$, $``P~wave"$, and $``D~wave"$ indicate that only that wave contributes to the width. In order to distinguish whether the wave comes from the initial state or the final state, we mark the final partial waves with a prime symbol.
\begin{table}[!htb]
\caption{The contribution of different partial waves to the strong decay width (keV) of $D^{*}(2007)^{0}\rightarrow D^{0}\pi^{0}$.}
\begin{tabular}{c|cccc}
\diagbox{$0^{-}$}{$1^{-}$}&$~complete~$ & $~S~wave$ & $~P~wave$ &$~D~wave$ \\
 \hline
 $~complete'~$ & ${34.6}$ & ${72.0}$ & ${63.5}$ & ${7.49\times10^{-3}}$\\
\hline
 $S'~wave~$ & ${64.5}$  & ${91.0}$ & ${2.28}$ & ${2.77\times10^{-6}}$ \\
\hline
 $P'~wave~$ & ${4.62}$ & ${1.11}$ & ${1.02}$  & ${7.20\times10^{-3}}$ \\
 \hline
\end{tabular}
\label{II}
\end{table}

As can be seen from Table \ref{II}, for the strong decay $D^{*}(2007)^{0}\rightarrow D^{0}\pi^{0}$, our relativistic result is {34.6} keV, but the non-relativistic contribution, i.e. the $S\times S'$  term in the table, is {91.0} keV, much larger than {34.6} keV, indicating that the relativistic effect is significant in this process. Although the wave function of a vector meson contains an $S$-wave, $P$-wave, and $D$-wave, as can be seen from Table \ref{II}, the contribution of the $D$-wave is negligible. Thus, we can conclude that the wave functions of a vector meson and a pseudoscalar meson are composed of both a non-relativistic $S$-wave and a relativistic $P$-wave.

\begin{table}[!htb]
\caption{The contribution of different partial waves to the strong decay width (keV) of $D^{*}(2010)^{+}\rightarrow D^{0}\pi^{+}$.}
\begin{tabular}{c|cccc}
\hline
 \diagbox{$0^{-}$}{$1^{-}$}&$~complete~$ & $~S~wave$ & $~P~wave$ &$~D~wave$ \\
 \hline
 $~complete'~$ & ${55.8}$ & ${117}$ & ${10.4}$ & ${9.80\times10^{-3}}$\\
\hline
 $S'~wave~$ & ${102}$  & ${143}$ & ${3.39}$ & ${4.66\times10^{-6}}$ \\
\hline
 $P'~wave~$ & ${6.98}$ & ${1.36}$ & ${1.91}$  & ${9.37\times10^{-3}}$ \\
 \hline
\end{tabular}
\label{III}
\end{table}

In the $D^{*}(2010)^{+}\rightarrow D^{0}\pi^{+}$ and $D^{*}(2010)^{+}\rightarrow D^{+}\pi^{0}$ decay processes, the contributions of the partial waves to the widths are similar to each other, and also similar to that of $D^{*}(2007)^{0}\rightarrow D^{0}\pi^{0}$. Thus, we only show the result for $D^{*}(2010)^{+}\rightarrow D^{0}\pi^{+}$ in Table \ref{III}. The main contributions of the initial state $D^{*}(2010)^{+}$ and the final state $D^{0}$ all come from the $S$-waves. The non-relativistic result is {143} keV, and the relativistic correction is also significant, offsetting most of the non-relativistic result, resulting in a width of {55.8} keV.

\subsection{EM decay widths of $D^{*}(2007)^{0}$ and $D^{*}(2010)^{+}$}

The EM decay widths of $D^{*}(2007)^{0}\rightarrow D^{0}\gamma$ and $D^{*}(2010)^{+}\rightarrow D^{+}\gamma$ are calculated as
\begin{eqnarray}\label{am33}
\Gamma(D^{*}(2007)^{0}\rightarrow D^{0}\gamma)={19.4~\rm{keV}},
\end{eqnarray}
\begin{eqnarray}\label{am34}
\Gamma(D^{*}(2010)^{+}\rightarrow D^{+}\gamma)={0.84~\rm{keV}}.
\end{eqnarray}

\begin{table}[!htb]
\caption{The EM decay widths (keV) of ${D^*}{(2007)^0}$ and ${D^*}{(2010)^+}$.}
\label{IV}
\begin{tabular}{c|cc ccc ccc}
\hline
&  ~~ {\cite{9610412v1(1)}}~~ & ~~{\cite{close2005}}~~ & ~~{\cite{aliev1996}} ~~ & ~~{\cite{jaus1996}} ~~ & ~~{\cite{kher2017}} ~~ & ~~{\cite{devlani2013}} ~~ & ~~ours ~~ &~~Ex\cite{PDG} ~~\\
\hline
\hline
$\Gamma(D^{*}(2007)^{0}\rightarrow D^{0}\gamma)$ & $ 12.9\pm2 $ & $ 32 $ & $ 14.4 $ &21.69 & & & $ {19.4} $ &  \\
\hline
$\Gamma(D^{*}(2010)^{+}\rightarrow D^{+}\gamma)$ & $ 0.23\pm0.1 $ & $ 1.8 $ & $ 1.5 $ & $ 0.56 $ & $ 0.271 $ & $ 0.339 $ & $ {0.84} $ & $ 1.33\pm0.4 $ \\
 \hline
\end{tabular}
\end{table}
As shown in Table \ref{IV}, our result for decay $\Gamma(D^{*}(2007)^{0}\rightarrow D^{0}\gamma)$ is in good agreement with the prediction in {Ref. \cite{jaus1996}} and close to the results of {Refs. \cite{9610412v1(1),aliev1996}}, but significantly smaller than the theoretical result of Ref. \cite{close2005}. {If we ignore the decay channel $D^{*}(2007)^{0}\to D^0 e^+e^-$, whose branching ratio $(3.91\pm 0.33)\times 10^{-3}$ is very small, we estimate the total width of $D^{*}(2007)^{0}$ to be
\begin{eqnarray}
&\Gamma(D^{*}(2007)^{0})\simeq \Gamma(D^{*}(2007)^{0}\rightarrow D^{0}\pi^{0})\nonumber\\
&\quad+\Gamma(D^{*}(2007)^{0}\rightarrow D^{0}\gamma)={54.0~\rm{keV}}.
\end{eqnarray}
Our estimate is consistent with the predictions of 53$\pm5\pm7$ \rm{keV} in Ref.\cite{1210.5410v2}, 55.9$\pm$1.6 \rm{keV} in Ref.\cite{rosner2013} and 59.6$\pm$1.2 \rm{keV} in Ref.\cite{cheung2014}, but  smaller than the 65.09 \rm{keV} in Ref.\cite{jaus1996} and 68$\pm$17 \rm{keV} in Ref.\cite{10.1140@epjc@s10052}.
Our results for the branching ratios $Br(D^{*0}\to {D^0}{\pi^0})=64.1\%$ and $Br(D^{*0}\rightarrow{D^0}{\gamma})=35.9\%$ are in good agreement with the experimental data of $64.7\pm0.9\%$ and $35.3\pm0.9\%$ from PDG, respectively.}

For $D^{*}(2010)^{+}$, our prediction of  $\Gamma(D^{*}(2010)^{+}\rightarrow D^{+}\gamma)={0.84}$~keV  is close to the experimental data from PDG \cite{PDG} and {larger than the results of Refs.\cite{9610412v1(1),jaus1996,kher2017,devlani2013}}, all of which are significantly smaller than the theoretical predictions of Refs. \cite{close2005,aliev1996}, indicating that more careful studies on this topic are still needed.

Since the wave functions of $D^{*}(2007)^{0}$ and $D^{*}(2010)^{+}$, as well as $D^{0}$ and $D^{+}$, are approximately equal, the significant difference between the decay widths $\Gamma(D^{*}(2007)^{0}\rightarrow D^{0}\gamma)$ and $D^{*}(2010)^{+}\rightarrow D^{+}\gamma$ mainly comes from the different quark charges inside the meson. In the transition amplitude, Eq.(\ref{am9}), $Q_{1}=\frac{2}{3}e$ and $Q_{2}=-\frac{2}{3}e$ for the transition $D^{*}(2007)^{0}\rightarrow D^{0}$, and $Q_{1}=\frac{2}{3}e$ and $Q_{2}=\frac{1}{3}e$ for $D^{*}(2010)^{+}\rightarrow D^{+}$. Therefore, for the $D^{*}(2007)^{0}\rightarrow D^{0}\gamma$ decay process, the contributions of the two graphs in Fig. \ref{feynman} are additive, while for $D^{*}(2010)^{+}\rightarrow D^{+}\gamma$, the contributions of the two graphs are subtractive, resulting in a significant difference in the widths of the these two processes.

To investigate the contribution of relativistic corrections in electromagnetic decays, we present the contributions of different partial waves in the wave function to the decay width, and the results are shown in Tables \ref{V} and \ref{VI}. As can be seen from Table \ref{V}, the non-relativistic contribution is {2.60} keV in the process of $D^{*}(2007)^{0}\rightarrow D^{0}\gamma$, which is much smaller than the final result of {19.4} keV, indicating that in this decay, the relativistic correction contributes the most.

\begin{table}[!htb]
\caption{The EM decay width (keV) of different partial waves for $D^{*}(2007)^{0}\rightarrow D^{0}\gamma$.}
\begin{tabular}{c|cccc}
\hline
 \diagbox{$0^{-}$}{$1^{-}$}&$~complete~$ & $~S~wave$ & $~P~wave$ &$~D~wave$ \\
 \hline
 $~complete'~$ & ${19.4}$ & ${9.54}$ & ${1.75}$ & ${1.83\times10^{-5}}$\\
\hline
 $S'~wave~$ & ${7.59}$  & ${2.60}$ & ${1.46}$ & ${4.20\times10^{-3}}$ \\
\hline
 $P'~wave~$ & ${2.73}$ & ${2.18}$ & ${0.0136}$  & ${3.67\times10^{-3}}$ \\
 \hline
\end{tabular}
\label{V}
\end{table}

Table \ref{VI} shows the results for $D^{*}(2010)^{+}\rightarrow D^{+}\gamma$. Similarly, the non-relativistic $S\times S'$ term gives a contribution of {0.135} keV, which is significantly smaller than the final result of {0.840} keV and also smaller than the contributions of the $S\times P'$ and $P\times S'$ terms, which are relativistic.
The conclusion that the non-relativistic contribution does not dominate is not uncommon. Unlike the electromagnetic process dominated by the $E1$ transition, which is $E1+M2+E3$ decay in our calculation \cite{liwei,du}, the $E1$ decay generally contributes the most. This $D^{*}\rightarrow D\gamma$ process  is $M1+E2+M3+E4$ decay in our method \cite{liwei,du}. The leading-order contribution comes from the non-relativistic $S\times S'$ term, which is the $M1$ transition, whose contribution is usually not the dominant one, and is usually comparable to or smaller than the contribution of the next-order $E2$ transition, from the $S\times P'$ and $P\times S'$ terms. $P\times P'$ and $D\times S'$ are the $M3$ transitions, and $D\times P'$ is the $E4$ transition, both of which contribute relatively little.

\begin{table}[htp]
\caption{The EM decay width (keV) of different partial waves for $D^{*}(2010)^{+}\rightarrow D^{+}\gamma$.}\label{VI}
\begin{tabular}{c|cccc}
\hline
 \diagbox{$0^{-}$}{$1^{-}$}&$~complete~$ & $~S~wave$ & $~P~wave$ &$~D~wave$ \\
 \hline
 $~complete'~$ & ${0.840}$ & ${0.193}$ & ${0.221}$ & ${5.42\times10^{-5}}$\\
\hline
 $S'~wave~$ & ${0.0503}$  & ${0.135}$ & ${0.331}$ & ${2.43\times10^{-4}}$ \\
\hline
 $P'~wave~$ & ${0.480}$ & ${0.650}$ & ${0.0111}$  & ${6.78\times10^{-5}}$ \\
 \hline
\end{tabular}
\end{table}

\section{Discussion and conclusion}

In this article, we used the Salpeter equation method to study the strong decays and radiative EM decays of $D^{*}(2007)^{0}$ and $D^{*}(2010)^{+}$.
{ Our results for $D^{*}(2010)^{+}$ strong decays are in good agreement with experimental data. For $D^{*}(2007)^{0}$, we obtain $\Gamma(D^{*}(2007)^{0}\to D^{0}\pi^{0})=34.6~\rm{keV}$, $\Gamma(D^{*}(2007)^{0}\rightarrow D^{0}\gamma)=19.4~\rm{keV}$, and then its full width is estimated as $\Gamma(D^{*}(2007)^{0})=54.0~\rm{keV}$. Although the $D^{*}(2007)^{0}$ total width is not available experimentally, the branching ratios of $D^{*}(2007)^{0}\to D^{0}\pi^{0}$ and $D^{*}(2007)^{0}\rightarrow D^{0}\gamma$ we obtained agree well with experimental data.}

The emphasis of this paper is on the relativistic corrections. In our method, the relativistic wave function of pseudoscalar $D$ meson is not a pure $S$-wave, but a mixture of the non-relativistic $S$-wave and relativistic $P$-wave, while the wave function of the vector $D^*$ meson includes both the non-relativistic $S$-wave and the relativistic $P$-wave and $D$-wave. Therefore, in this method, the radiative EM decay ${D^{*}\rightarrow{D}\gamma}$ is not a non-relativistic $M1$ transition, but an $M1+E2+M3+E4$ decay. We also note that in the process of strong decay $D^{*}\rightarrow{D}{\pi}$, the non-relativistic contribution is dominant, while in EM decay ${D^{*}\rightarrow{D}\gamma}$, the relativistic correction is dominant.

{\bf Acknowledgments}
This work was supported by the National Natural Science Foundation of China (NSFC) under Grant No. 12075073. W. Li is also supported by Hebei Agricultural University introduced talent research special project (No. YJ2024038).


\begin{thebibliography}{50}
\vspace{3mm}
\bibitem{1976}I. Peruzzi, \emph{et al}., Phys. Rev. Lett. \textbf{37}, 569 (1976).
\bibitem{1977}G. J. Feldman, \emph{et al}., (Mark I Collaboration), Phys. Rev. Lett. \textbf{38}, 1313 (1977).
\bibitem{slac-pub-1973(1)}G. Goldhaber, \emph{et al}., (Mark I Collaboration), Phys. Lett. B \textbf{69}, 503 (1977).
\bibitem{SLAC-PUB-2916}M. W. Coles, \emph{et al}., Phys. Rev. D \textbf{26}, 2190 (1982).
\bibitem{JADE Collaboration}W. Bartel \emph{et al}., (JADE Collaboration), Phys. Lett. B \textbf{161}, 197 (1985).
\bibitem{HRS}E. H. Low \emph{et al}., Phys. Lett. B \textbf{183}, 232 (1987).
\bibitem{SLAC-PUB-4518}J. Adler, \emph{et al}., (Mark III Collaboration), Phys. Lett. B \textbf{208}, 152 (1988) (Phys. Lett. B \textbf{227}, 501 (1989) (erratum)).
\bibitem{0000089}S. Abachi \emph{et al}., Phys. Lett. B \textbf{212}, 533 (1988).
\bibitem{ppe-92-017}S. Barlag \emph{et al}., (ACCMOR Collaboration), Phys. Lett. B \textbf{278}, 480 (1992).
\bibitem{butler1992}F. Butler \emph{et al}., (CLEO Collaboration), Phys. Rev. Lett. \textbf{69}, 2041 (1992).
\bibitem{albrecht1995}H. Albrecht \emph{et al}., (ARGUS Collaboration), Z.Phys.C \textbf{66}, 63 (1995).
\bibitem{bartelt1998}J. Bartelt \emph{et al}., (CLEO Collaboration), Phys. Rev. Lett. \textbf{80}, 3919 (1998).
\bibitem{PDG}S. Navas \emph{et al}., (Particle Data Group), Phys. Rev. D \textbf{110}, 3, 030001 (2024).
\bibitem{Print-88-0167}G. A. Miller, P. Singer, Phys. Rev. D \textbf{37}, 2564 (1988).
\bibitem{eichten1980}E. Eichten, K. Gottfried, T. Kinoshita, K. D. Lane, T. M. Yan, Phys. Rev. D \textbf{21}, 203 (1980).
\bibitem{pham1982}T. N. Pham, Phys. Rev. D \textbf{25}, 2955 (1982).
\bibitem{SLAC-PUB-3522}R. L. Thews, A. N. Kamal, Phys. Rev. D \textbf{32}, 810 (1985).
\bibitem{9203137}L. Angelos, G. P. Lepage, Phys. Rev. D \textbf{45}, R3021 (1992).
\bibitem{9209239v1}P. L. Cho, H. Georgi, Phys. Lett. B \textbf{296}, 408 (1992) (Phys. Lett. B \textbf{300}, 410 (1993) (erratum)).
\bibitem{9610412v1(1)}S.-L. Zhu, W.-Y. P. Hwang, Z.-S. Yang, Mod. Phys. Lett. A \textbf{12}, 3027 (1997).
\bibitem{9209262v1(1)}H.-Y. Cheng, C.-Y. Cheung, G.-L. Lin, Y.-C. Lin, T.-M. Yan, H.-L. Yu, Phys. Rev. D \textbf{47}, 1030 (1993).
\bibitem{9406300v1(1)}P. J. O'Donnell, Q. P. Xu, Phys. Lett. B \textbf{336}, 113 (1994).
\bibitem{aliev1996}T. M. Aliev, D. A. Demir, E. Iltan, N. K. Pak, Phys. Rev. D \textbf{54}, 857 (1996).
\bibitem{eichten}M. Di Pierro, E. Eichten, Phys. Rev. D \textbf{64}, 114004 (2001).
\bibitem{close2005}F. E. Close, E. S. Swanson, Phys. Rev. D \textbf{72}, 094004 (2005).
\bibitem{salpeter1951}E. E. Salpeter, H. A. Bethe, Phys. Rev. \textbf{84}, 1232 (1951).
\bibitem{salpeter1952}E. E. Salpeter, Phys. Rev. \textbf{87}, 328 (1952).
\bibitem{1305.1067v2}T.-H. Wang, G.-L. Wang, H.-F. Fu, W.-L. Ju, JHEP \textbf{07}, 120 (2013).
\bibitem{tan2018}X.-Z. Tan, T. Wang, Y. Jiang, S.-C. Li, Q. Li, G.-L. Wang, C.-H. Chao, Eur. Phys. J. C \textbf{78}, 583 (2018).
\bibitem{wang-2018}Z.-H. Wang, Y. Zhang, T.-H. Wang, Y. Jiang,  Q. Li, G.-L. Wang, Chin. Phys. C \textbf{42}, 123101 (2018).
\bibitem{main}T.-H. Wang, G.-L. Wang,  Phys. Lett. B \textbf{697}, 233 (2011).
\bibitem{wang2013}T.-H. Wang, G.-L. Wang, Y. Jiang, W.-L. Ju, J. Phys. G \textbf{40}, 035003 (2013).
\bibitem{PhysRevD.108.033003}S.-Y. Pei, W. Li, T.-T. Liu, M. Han, G.-L. Wang, T.-H. Wang, Phys. Rev. D \textbf{108}, 033003 (2023).
\bibitem{mandelstam1955}S. Mandelstam, Proc. R. Soc. Lond. A \textbf{233}, 248 (1955).
\bibitem{0505205v2}C.-H. Chang, C. S. Kim, G.-L. Wang, Phys. Lett. B \textbf{623}, 218 (2005).
\bibitem{mainext}C. S. Kim, G.-L. Wang, Phys. Lett. B \textbf{584}, 285 (2004). Phys. Lett. B \textbf{634}, 564 (2006) (erratum).
\bibitem{chao-hsi2006}C.-H. Chang, J.-K. Chen, G.-L. Wang, Commun. Theor. Phys. \textbf{46}, 467 (2006).
\bibitem{0512009v1}G.-L. Wang, Phys. Lett. B \textbf{633}, 492 (2006).
\bibitem{JHEP05(2022)006}G.-L. Wang, T.-H. Wang, Q. Li, C.-H. Chang, JHEP \textbf{05}, 006 (2022).

\bibitem{liwei2}W. Li, S.-Y. Pei, T. Wang, T.-F. Feng, G.-L. Wang, Phys. Rev. D \textbf{109}, 036011 (2024).
\bibitem{pei}S.-Y. Pei, W. Li, T. Wang, G.-L. Wang, JHEP \textbf{08}, 191 (2024).


\bibitem{chang2010}C.-H. Chang, G.-L. Wang, Sci. China Phys. Mech. Astron. \textbf{53}, 2005 (2010).
\bibitem{cheung2014}C.-Y. Cheung, C.-W. Hwang, JHEP \textbf{04}, 177 (2014).
\bibitem{jaus1996}W. Jaus, Phys. Rev. D \textbf{53}, 1349 (1996). Phys. Rev. D \textbf{54}, 5904 (1996) (erratum).



\bibitem{kher2017}V. H. Kher, A. K. Rai, J. Phys. Conf. Ser. \textbf{934}, 1, 012036 (2017).
\bibitem{devlani2013}N. Devlani, A. K. Rai, Int. J. Theor. Phys. \textbf{52}, 2196 (2013).




\bibitem{1210.5410v2}D. Becirevic, F. Sanfilippo, Phys. Lett. B \textbf{721}, 94 (2013).
\bibitem{rosner2013}J. L. Rosner, Phys. Rev. D \textbf{88}, 3, 034034 (2013).
\bibitem{10.1140epjc@s10052}D. Becirevic, B. Haas,  Eur. Phys. J. C \textbf{71}, 1734 (2011).

\bibitem{liwei}W. Li, S.-Y. Pei, T. Wang, Y.-L. Wang, T.-F. Feng, G.-L. Wang, Phys. Rev. D \textbf{107}, 113002 (2023).
\bibitem{du}X.-Y. Du, S.-Y. Pei, W. Li, M. Jia, Q. Li, T. Wang, B. Wang, G.-L. Wang, Phys. Rev. D \textbf{110}, 113009 (2024).



\end{thebibliography}
\end{document}